# Enhancing Security Testing Software for Systems that Cannot be Subjected to the Risks of Penetration Testing Through the Incorporation of Multi-threading and and Other Capabilities

Matthew Tassava, Cameron Kolodjski, Jordan Milbrath, Jeremy Straub
Institute for Cyber Security Education and Research
North Dakota State University
1320 Albrecht Blvd., Room 258
Fargo, ND 58108
Phone: +1-701-231-8196
Fax: +1-701-231-8255
Email: matthew.tassava@ndsu.edu, cameron.kolodjski@ndsu.edu, jordan.milbrath@ndsu.edu
jeremy.straub@ndsu.edu

**Abstract**

The development of a system vulnerability analysis tool (SVAT) for complex mission critical systems (CMCS) produced the software for operation and network attack results review (SONARR). This software builds upon the Blackboard Architecture and uses its a rule-fact logic to assess model networks to identify potential pathways that an attacker might take through them via the exploitation of vulnerabilities within the network. The SONARR objects and algorithm were developed previously; however, performance was insufficient for analyzing large networks. This paper describes and analyzes the performance of a multi-threaded SONARR algorithm and other enhancements which were developed to increase SONARR's performance and facilitate the analysis of large networks.

## 1. Introduction

The software for operations and network attack results review (SONARR) [1] was designed to analyze the security of systems that cannot undergo traditional penetration testing due to the risks it poses. It is designed to identify vulnerabilities so that they can be remedied. SONARR builds on the rule-fact-based Blackboard Architecture to provide a modular expert system that analysts can employ to perform vulnerability assessment of complex mission critical systems (CMCS) that cannot be taken offline or risk disablement. The rule-fact nature of the algorithm facilitates variable granularity of system analysis and makes the reasoning of its decision making readily human-understandable.

SONARR has gone through several iterations of development and refinement [1,2]. Initial logic and algorithm development, that extended the rule-fact Blackboard Architecture, was performed. This included the design and development of key elements, such as the single-threaded SONARR algorithm, the objects that are used to model networks, and a rule-fact logic expansion mechanism to facilitate generalization within the Blackboard Architecture [1,3]. Second, an additional rule type, that resembles the rules used in the traditional Blackboard Architecture, was added. This rule type can assess facts anywhere in the network and facilitates the consideration of environmental variables and other information that is not related to a particular system or network connection. Additional objects, such as actions [4], were also added, amongst a number of other improvements, such as interface changes and usability improvements [2,5].

This paper builds on this prior work. It presents and assesses a new SONARR algorithm that incorporates multi-threading to facilitate the rapid processing of larger networks. Several other enhancements to facilitate working with larger networks are also discussed.

## 2. Background

SONARR is a network analysis tool [1] that builds upon the Blackboard Architecture to make logical assessments. The Blackboard Architecture is an expansion of expert systems which was introduced by Hayes-Roth [6]. It is based on technology in the HEARSAY-II system, which was developed for a DARPA-sponsored challenge to create a speech recognition system [7].

The Blackboard Architecture has demonstrated its effectiveness across a variety of uses, such as software testing [8], medical imaging [9,10], agent control in strategy games [11], robotic development [12-15], counter-terrorism [16], and mathematical proof creation [17]. The rule-fact logic employed by expert systems and the Blackboard Architecture have a natural interpretability to them, making it easier for humans to understand what the system is doing and why it reaches certain conclusions.

The explainability of artificial intelligence (AI) has become an important topic due to the growth in usage of neural networks and other opaque machine learning approaches [18,19]. Making an AI implementation explainable requires the AI's internal functions and decision-making to be easy to understand by humans [20-22]; however, a gap still exists in the definition of AI "explainability" [23].

Some eXplainable AI (XAI) techniques use a post-hoc explanation approach [24], which generates an explanation after a conclusion has been reached. This approach is unsuitable for many applications, as the AI should be able to explain its actual decisions and undergo auditing [25]. In [25], Ehsan, et al. explore expanding AI explainability to parameters outside the algorithm itself using social transparency to provide explanations with historical and contextual qualities. Their study participants indicated that context-oriented output is needed for user trust in AI systems.

One area that can benefit from trustworthy artificial intelligence is cybersecurity assessment. Existing tools already perform tasks such as dynamic malware analysis [26], intrusion detection [27,28], and log analysis [29-31]. These tools help increase efficiency and human productivity; however, as systems grow in complexity, the number of areas that require human attention and understanding grows. An explainable artificial intelligence, that can learn from incidents and vulnerability disclosures to detect and respond to threats beyond its specific training, is a key area of need.

## 3. Network Objects

This section describes the objects which are used by SONARR to model networks for analysis. SONARR uses these model networks, which are manually created by analysts, as digital twins or cousins to analyze their security. The networks can have the abstraction level desired by their creator, allowing them to be as detailed or abstract as needed.

Containers are entities within these networks (e.g., routers, computers, and printers) and links are used to model their interconnections, creating logical relationships that SONARR can assess. Common properties and facts are used to store the attributes of entities and the environment that can be leveraged or manipulated (e.g., identifiers, configurations, and power states). Rules, which come in the two forms 'normal' and 'generic', are used to model attack techniques. Using these objects, SONARR can analyze the network as a whole and on a specific container-link-container level.

Each type of object within SONARR is now discussed.

### 3.1. Organizational Objects

This section describes the objects used to model the components of a network SONARR analysis. These objects provide the network that the SONARR algorithm will traverse during processing. They allow analysts to follow the logical processes on a component-by-component basis.

A model network is comprised of two entity objects: containers and links. These objects are used to represent the components found in a network (or any web of connections that the user wants to represent and analyze). They can be used to model software, hardware, people, places, or even events (for certain uses). They are used to embody the user's understanding of the network. Each entity houses a collection of facts that are unique, which are used by the SONARR algorithm's assessment logic. Figures 1 and 2 depict a container and link, respectively.

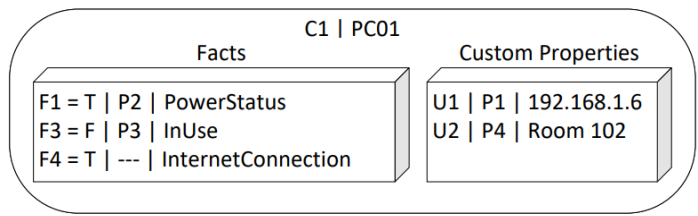

Figure 1. Depiction of a container.

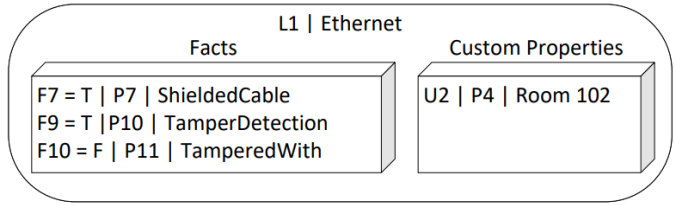

Figure 2. Depiction of a link.

Containers and links house objects that store data. These objects are facts and custom properties.

Facts store a value that is associated with the particular container. If a fact stores a type of information that is relevant to multiple containers, it can be associated with a common property to embody this. A fact can be associated with a maximum of one common property.

Custom properties are objects that allow the storage of extra description information for any entity or fact. They are not currently assessed; however, they are intended to be a part of the algorithm's logical assessment in the future. A custom property is depicted in Figure 3.

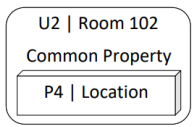

Figure 3. Depiction of a custom property.

Facts, common properties and custom properties are discussed in more detail in the subsequent subsection.

### 3.2. Logical Assessment Objects

This section describes the objects used by the SONARR algorithm for its rule-fact logic assessments.

#### 3.2.1. Facts

Facts (shown in Figure 4) are objects that hold Boolean values related to the attributes of an entity or the network environment. They are unique to the entity that they belong to, and identified by their ID values. Two or more entities can have a fact with the same property, but those facts would not have the same ID. If a fact represents an attribute outside a specific entity or relevant to the network as a whole, it is considered an environment fact and stored within the network itself. Many facts represent the same thing within their respective entities, but have different values. However, they require identical treatment within the SONARR algorithm's logic. The common property object was created for this purpose.

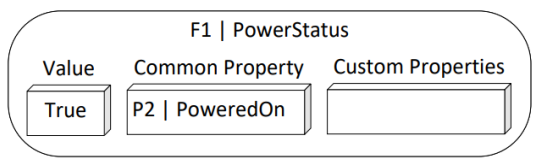

Figure 4. Depiction of a fact.

Common properties are objects designed to create an association between facts, effectively giving the facts an information type. This association allows generic rules to treat the common property-associated facts as being the same across multiple objects. Because of this, rules can be created that can be used across every connection in a network that has facts satisfying their pre- and postconditions. Figure 5 depicts a common property. Rules are discussed in the next subsection.

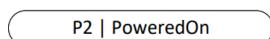

Figure 5. Depiction of a common property.

*3.2.2. Rules*

SONARR uses two types of rules: generic rules and normal rules. Each is now discussed.

Generic rules are designed to be applicable across any connection in the network, as long as the connection entities have facts that satisfy the rule conditions. One generic rule can, thus, be created and be used effectively across hundreds of connections in the network. Without this generality, a user would have to create the same normal rule hundreds of times with the same conditions associated with each set of unique facts.

Both rule objects have two sets of conditions: preconditions and postconditions. The preconditions for a generic rule are satisfied if all three entities in the connection it is assessing (two connected containers and a link) have the appropriate value/common property pairings. If any pairing is incorrect or missing, the rule does not apply to that connection and the postconditions would not be triggered. Postconditions-required common properties must also be present for a rule to be applied to a connection. Postconditions for generic rules also designate a value/common property pairing; however, the algorithm uses these to make changes to the connection entities' facts and store the resulting entities as variants a reality path to use in future processing by the algorithm. A generic rule is shown in Figure 6.

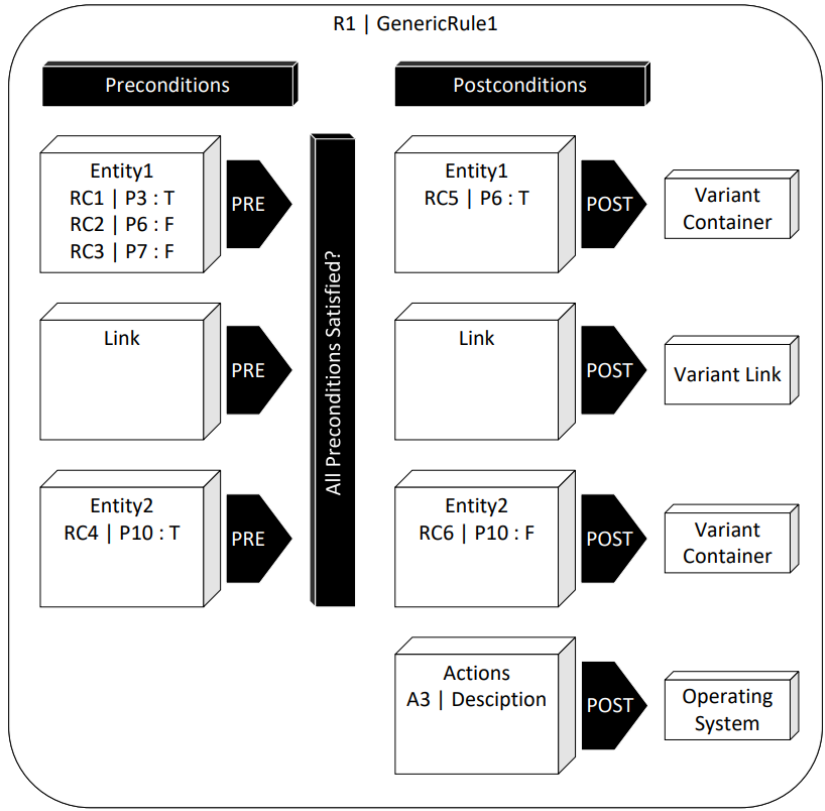

Figure 6. Depiction of a generic rule.

The variant objects, shown in Figure 6, are elements of SONARR's traversal processing, along with connections and reality paths. These objects are discussed in Section 3.3.

The second type of rule object is the normal rule. This rule type also utilizes pre- and postconditions. However, the preconditions for normal rules include only value/fact pairings and do not use common properties. The postconditions of a normal rule can alter fact values on either a fact or common property basis. A postcondition specifying a value/fact pairing will alter just that particular fact; however, a postcondition specifying a value/common property pairing will find all facts in the network associated with that common property and alter all of their values. Notably, a fact that is associated with a common property can serve as a pre- or post-condition for a normal rule, based on its individual ID.

Only normal rules can access and utilize environment facts for their logic. This makes them useful to make large-scale changes that affect the entire network. A normal rule is shown in Figure 7.

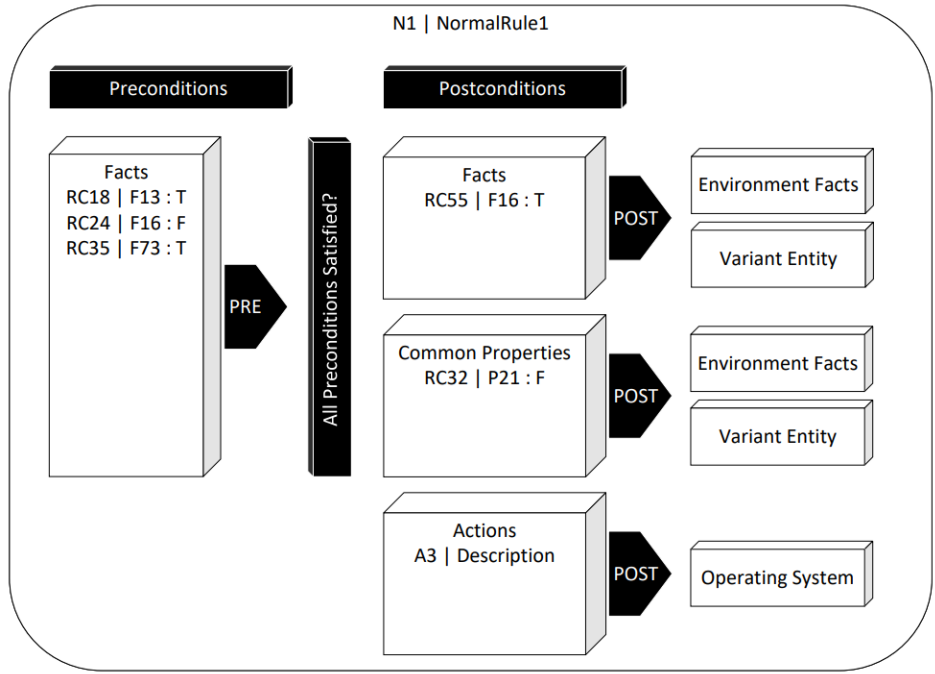

Figure 7. Depiction of a normal rule.

Actions are objects that can be run when a rule is triggered. They build on the core Blackboard Architecture concept of actualization and their specific implementation was presented in [4]. They allow the SONARR algorithm to issue commands that are executed on the host operating system. This allows environment manipulation, data collection and other beneficial activities to be performed during traversal processing. Figure 8 depicts an action object.

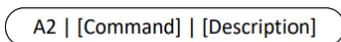

Figure 8. Depiction of an action.

### 3.3. Traversal Objects

This section describes the objects used by the SONARR algorithm to facilitate traversal processing.

#### 3.3.1. Variants

The term variant is used to denote entities and facts that have been accessed, whether a change is made to them or not, and stored in a reality path. Variants allow altered versions of entities to be stored in memory for other reality paths to use. This reduces the amount of memory used when the algorithm is building reality paths and stores the changes without altering the base network itself. Not altering the base network allows multiple reality paths to be built concurrently, using the same network, without rules that are triggered in one reality path altering the results in any others. A variant is depicted in Figure 9.

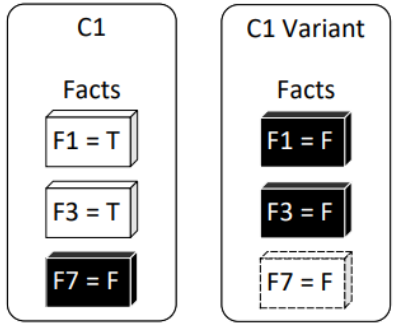

Figure 9. Depiction of fact utilization within a variant entity.

*3.3.2. Connections*

Connection objects temporarily model container-link-container relationships within the SONARR network. Connections are created by duplicating the relevant network containers and links during the traversal process. Rules are run using the copies, making the postcondition-specified alterations. The altered objects are then stored in the reality path, creating a history of the changes made along that path. Each connection stores a list of the rules triggered, allowing for better post-run analysis. A connection is depicted in Figure 10.

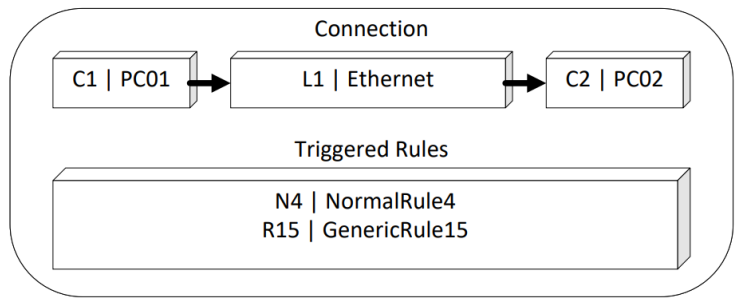

Figure 10. Depiction of a connection.

*3.3.3. Reality Paths*

Reality paths are the output object of SONARR. They are built connection-by-connection, cloning existing paths that new paths are built upon. Once the designated end container is reached, they are stored to be provided to the user as output. Each reality path contains a list of connections, which record the route taken through the network, along with any changes made. Reality paths make changes to the cloned entities, keeping every manipulation within the reality path itself, so that other reality paths can be built concurrently without interference. A reality path is depicted in Figure 11.

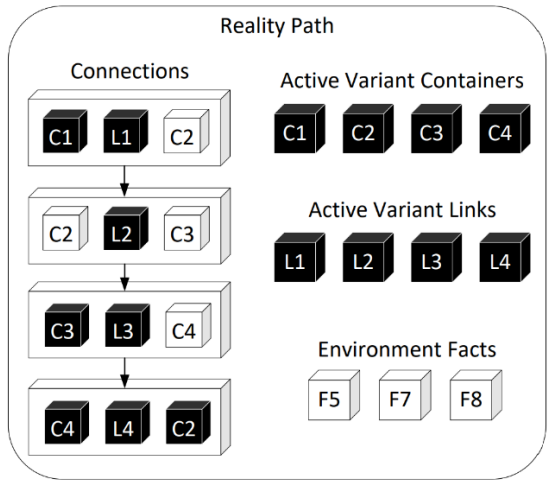

Figure 11. Depiction of a reality path.

## 4. SONARR Algorithm

The SONARR algorithm was developed to perform exhaustive enumerations of networks, typically in a depth-first pattern. It is designed to search until all paths to a goal are identified. Although it successfully achieves this goal, this can be a time-consuming process for larger networks. A faster implementation of the algorithm, thus, is beneficial. A key step toward this goal is to use multi-threading.

The following subsections discuss the single-threaded and multi-threaded implementations of the SONARR algorithm.

### 4.1. In-Memory Single-Threaded SONARR

This section describes the in-memory single-threaded SONARR algorithm. This algorithm uses an exhaustive depth-first search to find all possible paths through a given SONARR network, from a specified start container to a specified end container. The search algorithm utilizes two lists of reality paths to keep track of traversals: a list of in-progress paths and a list of final (start-to-end) paths.

In each iteration of the algorithm's main loop, a reality path is taken from the list of in-progress paths. The outgoing links from the end container of the last connection in this path are iterated through. For each of these outgoing links, a new path is created by cloning the selected path and assigning it a new ID. This effectively branches pathways, because each new path retains the same connection history. They differ in the new connection, under rule assessment, which is created using either the most recent container-link-container variants or, if variants have not been created yet, clones of the base network entities.

Once a new connection is created, rules are evaluated. A rule-running loop is used for this purpose. The loop evaluates, at most, one normal rule and one generic rule per iteration. A rule will not be re-evaluated if it has already been triggered on the connection. Rule assessment of the connection stops if the triggered rule limit has been reached. This limit is a user-defined value that dictates the maximum number of generic rules that can be triggered when assessing each connection.

The rule-running loop begins by evaluating normal rules. Normal rules check for their preconditions, first, in the reality path's normal facts list. Then, the reality path's active variant containers are checked. Next, the active variant links are assessed. Finally, the base network's containers and links are searched.

If a rule's preconditions are met, the rule is added to the connection's triggered rules list, and the postconditions are applied to the path's normal facts and active variants. If an active variant does not exist, a new variant is created, based on the original entity in the network, and the postconditions are applied to that. Any actions associated with the rule will also be run at this time. Finally, the entities of the connection are updated to reference the correct active variants, based on changes made by the rule.

After the normal rules' evaluation, generic rules are evaluated. Unlike normal rules, generic rules' preconditions only need to assess common properties within the entities of the connection. If the rule's preconditions are met, the generic rule is added to the triggered rules list and the postconditions are applied to the entities of the connection. The reality path's active variant lists are then updated with the connection's entities. Finally, actions associated with the triggered rule are run.

If either a normal or generic rule was triggered during an iteration, the loop repeats and rules are assessed again, with the fact changes applied by the newly triggered rules. If no rules are triggered during an iteration, the rule-running loop terminates.

After the completion of the rule-running loop, the connection is assessed using the path termination heuristic. This terminates a path if the new connection, produced by rule assessment, is already in the path history. If the connection passes this heuristic and at least one generic rule has triggered for the connection, then the path is added to the algorithm's list of in-progress reality paths for subsequent processing in a later main loop iteration.

When a reality path is being processed that has a last container that matches the user-specified end container, a cloned reality path is created, like usual; however, the connection created has no link or second entity. The new connection only contains a first entity. This allows any rules that affect only the first entity of a connection to be assessed and apply any applicable changes. After any rules are run, the path is added to the final paths list.

Figure 12 depicts the single-threaded SONARR algorithm.

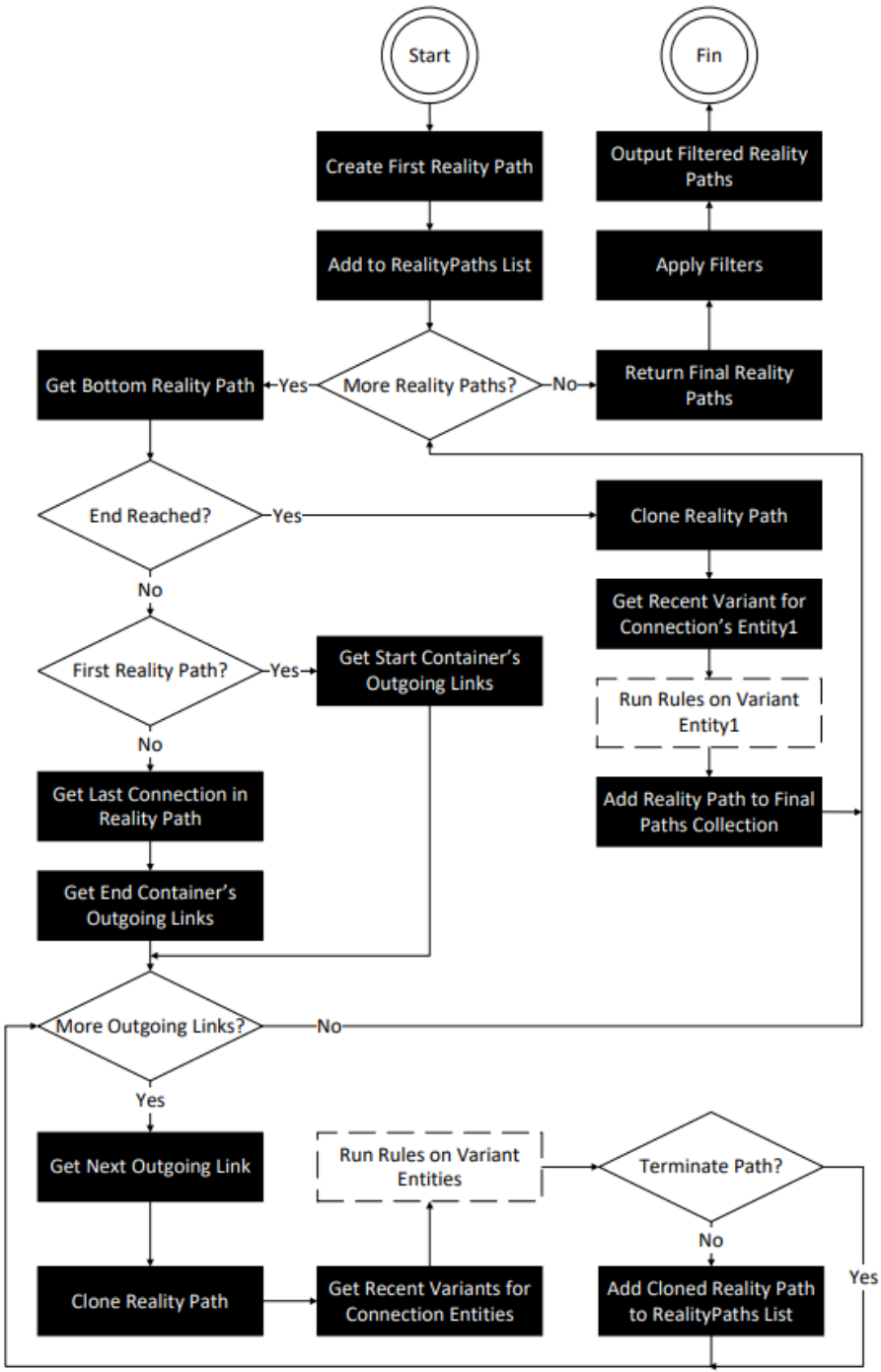

Figure 12. The in-memory single-threaded SONARR algorithm introduced in [2].

### *4.2. File-Writing Multi-Threaded SONARR Algorithm*

This section discuses a new SONARR algorithm implementation which is multi-threaded and writes final reality paths to a file as they are completed. Implementing these changes has allowed the algorithm to perform significantly faster than previous versions.

*4.2.1. Multi-Threaded SONARR Algorithm Overview*

The multi-threaded SONARR algorithm (shown in Figure 13) performs the same exhaustive search as the single-threaded algorithm, but with greater efficiency. This is due to two enhancements. First, this algorithm utilizes multiple threads by splitting the traversal algorithm's work into several concurrently running tasks. Second, it writes paths to disk instead of storing them in working memory, keeping more memory available for finding new paths. Each thread writes final paths to its own results file. When all tasks are complete, the files are merged into one final paths file. Each step in this algorithm is now discussed.

At the start of the algorithm's processing, the number of threads to use is determined based on the computer's processor count. A number of threads that is one less than the processor count is used, leaving one processor for the user interface (UI) and operating system (OS). A reality path stack is created for each thread to use as its working collection.

The first reality path is created based on the user-specified start container. Tasks are then created. The first task that is created receives the first reality path and begins traversing. The other threads simultaneously reach the reality path redistribution step (as shown in Figure 14), which performs the algorithm's load balancing. These threads have no paths to process, indicating they need to be given work. Thus, they wait for their assignments. When the first thread has ten queued paths, it gives five of them to the next available thread. As threads traverse and generate reality paths, they offload some of their reality path queue to idle threads waiting for work. They do this until every thread is processing. If a thread becomes idle, another thread will give half their workload to the idle thread and both will continue working. No duplication happens, because each reality path is unique by nature and they are transferred to other threads as a whole (i.e., no thread will have a reality path that is also in the working queue of another thread). Details about thread management and the load-balancing are presented in the next section.

The algorithm ends when the first thread sees that all threads (including itself) are available for work. This indicates that all paths have been processed. Overall, the algorithm is logically similar to the single-threaded algorithm; however, it has data transfer and storage changes that facilitate a multi-threaded processing workflow.

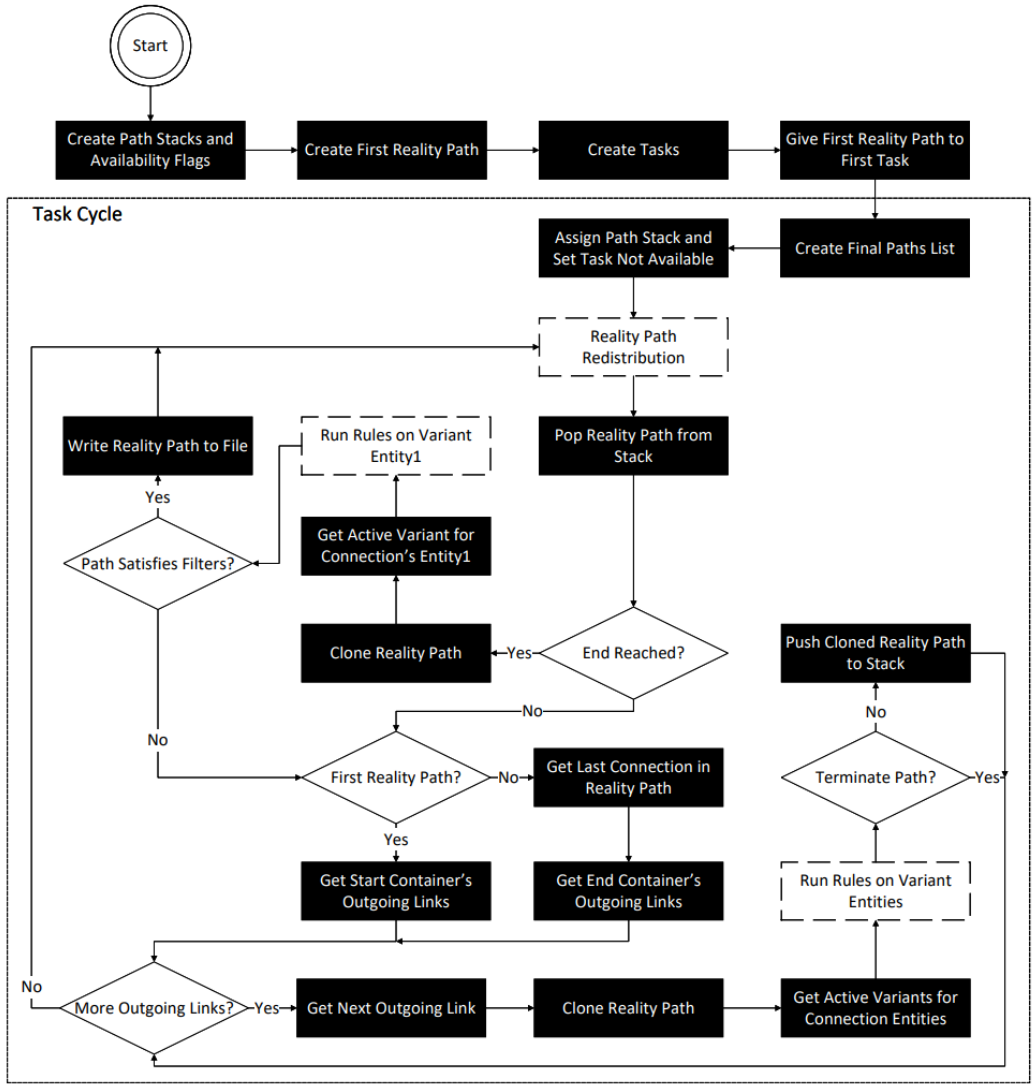

Figure 13. Multi-threaded algorithm flowchart.

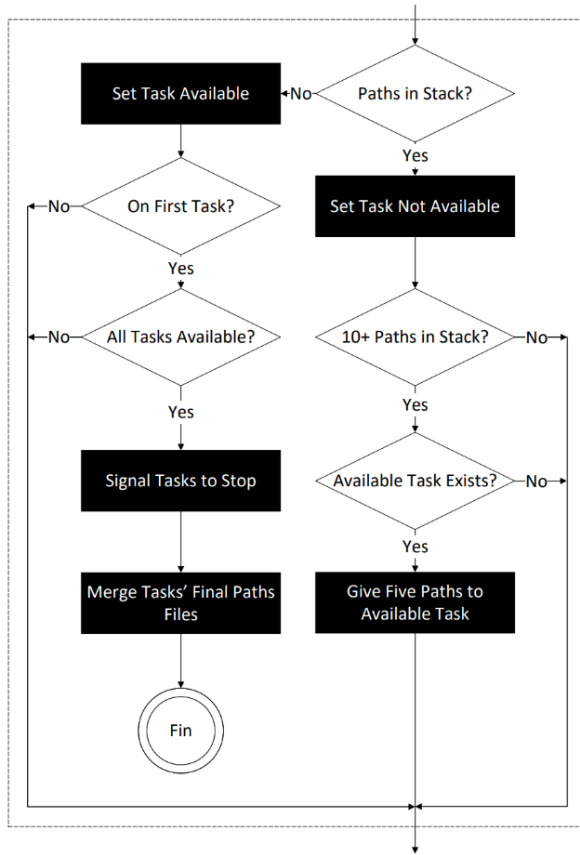

Figure 14. Reality path redistribution section of the multi-threaded algorithm.

*4.2.2. Tasks and Reality Path Management*

The multi-threaded SONARR algorithm uses a number of processing threads equal to the number of available processors in the system the application is running on minus one. This last processor is left for use by the UI and OS processes to ensure that the UI remains responsive to users during traversal.

Data for each of the processing threads is stored in arrays. Multiple stacks of reality paths are used to keep track of which paths still need to be examined by each thread. Each thread has its own stack that is pulled from each time it reads a new reality path to process. Using threads that create and use their own working stacks prevents cross-thread complications that could otherwise arise from resource sharing.

Reality paths vary in size, often to large degrees. This results in some threads accumulating queues of reality paths while others empty their stacks more quickly. To balance this load between the threads, whenever a thread has a quantity of pending reality paths that is larger than a setting value (which is ten by default), it checks to see if there are other threads that do not have paths to process. If a thread requires work, the thread with extra reality paths gives half of its workload to the idle thread. This ensures that threads spend as much time working as possible, and minimizes thread idle time.

Traversal is initiated by generating a reality path of the initial network at the start of the traversal, which is given to the first task to process. This process of redistributing tasks is, thus, also how threads, other than the first thread, receive initial reality paths to process. Only the first thread receives a reality path, initially. It processes this path until it reaches the redistribution threshold and then distributes paths to the rest of the tasks, each time it reaches the threshold. Since threads check to see if there are idle threads every processing iteration, the initial redistribution happens quickly.

The processing of each reality path is the same as with the single-threaded algorithm. Since each thread stores its own data, processing is not impeded by the locking of resources, except when a redistribution of reality paths between the tasks occurs. In large networks, with a large number of paths, redistribution does not happen frequently. This allows the threads to largely run independently of each other.

*4.2.3. File Writing and Storage*

Since a single file can only be modified by one thread at a time, each thread has its own set of files that it writes to during traversal. The set of files includes (and the titles of the files are):

- Availability
- Confidentiality
- Final paths
- ID
- Index
- Integrity
- Total run time
- Traversability chance

Each thread's instance of these files is named with the format "{File Title}-{Thread ID}.tmp". The final paths file stores information about each reality path itself. Each final reality path is written to the file in the order that they are found using the format shown in Figure 15. Note that some fields, indicated with a *, may have multiple instances up to the maximum 32-bit value.

| 0 | 4 | 8 |
|---|---|---|
| Reality Path ID | Number of Connections | |
| Connection Data* | | |
| Number of Environment Facts | Environment Fact Data* | |

Figure 15. Reality path storage data format.

Within the final paths file, connections are stored using the format shown in Figure 16. Connections store entities using the format shown in Figure 17. These entities store facts within them using the format shown in Figure 18.

| 0 | 4 | 8 |
|---|---|---|
| Connection ID | Entity 1 Data* or -1 if null | |
| Link Data* or -1 if null | Entity 2 Data* or -1 if null | |
| Number of Environment Facts | Environment Fact Data* | |

Figure 16. Connection storage data format.

| 0 | 4 | 8 |
|---|---|---|
| Entity ID | Number of Facts | |
| Fact Data* | | |

Figure 17. Entity storage data format.

| 0 | 4 | 5 |
|---|---|---|
| Fact ID | Value | |

Figure 18. Fact storage data format.

Storing reality path data in this way allows everything necessary about a reality path to be determined from the stored information and the network that was traversed to generate it. Each reality path entry can be read, in full, from its start position in the final paths file.

The index file is used to store the start positions of each reality path in the final paths file. Before each path's data is written to the final paths file, the position of the file's pointer is written to the index file. Each location is eight bytes long. This means that the a given path's position in the final paths file is found in the index file location n*8.

The remaining files are used to store data that is used to sort reality paths. Once each task completes its main traversal work, it reads each of its reality paths back into main memory from its index and final paths files, sorts them, and then writes the sorted data into the file corresponding to the method that was used to sort the reality paths.

The sorted data is written to each file in the formats shown in Figures 19 and 20. The sorting values that are stored as integers are written as four-byte values (as shown in Figure 19).

| 0 | 4 | 12 |
|---|---|---|
| Sort Value | Final Paths File Position | |

Figure 19. Sort value data format for four-byte values.

The sorting values that are stored as doubles are written as eight-byte values (as shown in Figure 20).

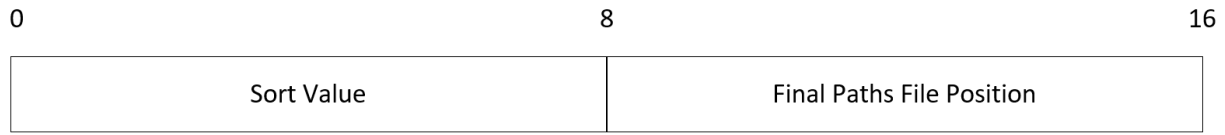

Figure 20. Sort value data format for eight-byte values.

When the threads complete their processing, the final paths and index files are merged. To do this, the data is transferred from the thread-specific files to a main file. Each thread stores positions in its index without regard to other threads. However, the location in the combined file is relative to the position of the start of the thread's data. For example, task four's first final path will be at the beginning of task four's final path file and index file. However, in the file that stores all of the final paths, it will no longer be at the beginning, since the paths from tasks one, two, and three will already have been written to that file.

For files beyond the first one, the pointers are updated. The position of the file pointer is noted after each thread is finished writing. Then, when the next thread's data is written to the file, that value is added to the pointer positions. These offsets are stored in an array to be used when sorting.

When a value is chosen to sort by, SONARR checks to see if a file exists for that sort method. If not, each of the threads' files for that sort method are merged to create a complete file for the specified sort method. This is done using a merge sort across files. A file stream is opened for each file, and the sort value is read into memory. The values from the files are compared, and the highest value is selected. The position is read from the file that had the highest sort value, and the offset array is consulted to determine the location of the corresponding reality path's data in the complete data file. Only the location of the reality path is written to the complete sort method file. Then, the file pointers for the rest of the streams are reset to their positions from before the sort value was read. Once the values from all of the streams have been written to the complete sort file, the streams are closed. Having combined files for each sort method means that if a method is selected again, the sort does not need to be performed again, saving time.

## 5. Experimentation

The experiments discussed in this paper focus on comparing the single-threaded and the multi-threaded SONARR algorithm implementations. Experimentation is also performed to assess the utility of a filtering mechanism.

The single-threaded algorithm was not run during these experiments. Instead, data collected for the single-threaded SONARR algorithm in [2] was used for the comparisons. Data for the multi-threaded algorithm was collected using the new SONARR software. Scenarios were selected from [2] and run using the multi-threaded algorithm that showed performance difficulty with the in-memory single-threaded implementation. These scenarios were tested with the multi-threaded algorithm to compare the performance between algorithm implementations.

The multi-threaded algorithm writes final paths into separate files and then merges and sorts these result files. The single-threaded algorithm does not. Therefore, for these tests, the merge/sort element of the software is disabled so that the multi-threaded and the single-threaded implementations are more directly compared.

Additionally, basic experimentation was done utilizing an upgraded filter mechanism. These filters are used during traversal processing, and are designed to allow a reality path to continue past the target container if any specified conditions are not met. The experimentation presented demonstrates the efficacy, effect of, and propositional logic behind filters, which may be useful for various use cases.

The tests presented were performed using a computer with an AMD Ryzen 7 5700U CPU and 8 GB of RAM.

*5.1. Compared Scenarios*

Six of the fifteen scenarios from [2] were used for testing the multi-threaded implementation, to compare results between threading types. The selected scenarios are shown in Table 1.

Table 1. Scenarios selected for multi-threaded algorithm testing.

|  | **Model 1** | **Model 2** | **Model 3** |
|---|---|---|---|
| **Scenario 1** |  |  |  |
| **Scenario 2** |  |  | X |
| **Scenario 3** |  | X | X |
| **Scenario 4** |  | X | X |
| **Scenario 5** |  |  | X |

In [2], scenarios 2, 4, and 5 for model 3 were stopped at certain thresholds. In the case of scenario 2, stopping was due to memory limitations due to the increasing number of final reality paths. In the case of scenarios 4 and 5, stopping was due to time constraints.

The tests performed for these scenarios with the multi-threaded SONARR implementation are similar to those performed in [2]. Two tests were performed for scenario 2: one that follows the protocol of the single-threaded test (stopping the algorithm when the final paths are above 1.2 million) and one that runs until completion. This allows for direct comparison between the algorithm versions and also builds upon the original test by collecting the results of a full run to completion. The time windows of scenarios 4 and 5 were increased, by 4 hours, to 24 hours. Figures 22 and 23 depict the models used for this testing: models 2 and 3, respectively.

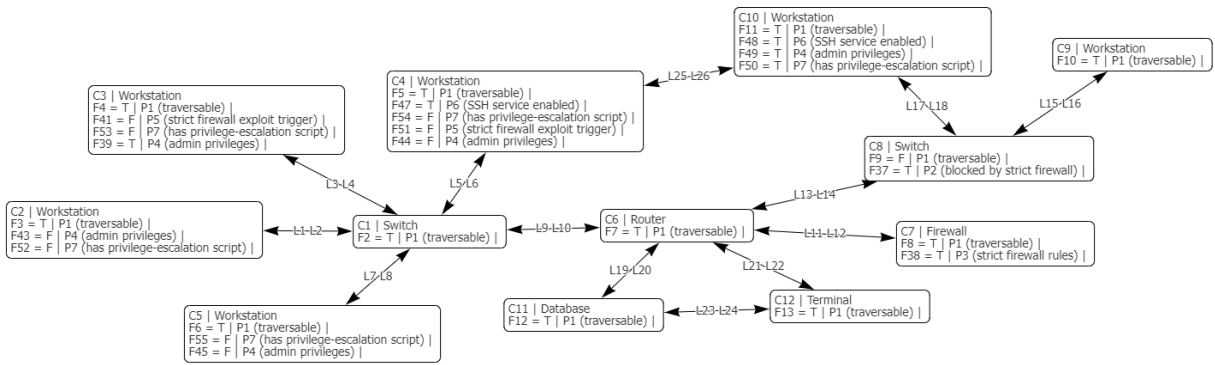

Figure 22. Model 2 visualization.

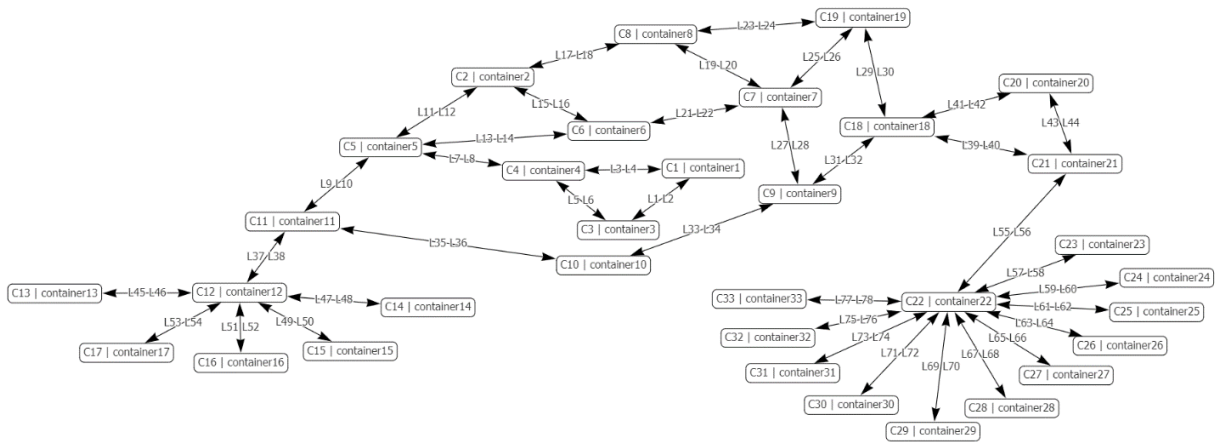

Figure 23. Model 3 visualization.

## 5.2. Filter mechanism

In addition to the new multi-threaded SONARR algorithm, this version of SONARR includes a filtering mechanism that allows for traversal through the end container, if the fact configuration of the end container does not match the filter's specifications.

To illustrate this new functionality, a simple model has been developed and imported into SONARR. This model is shown in Figure 24.

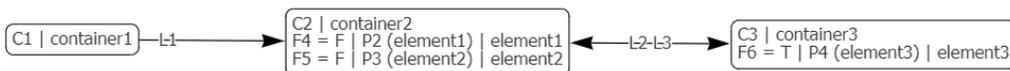

Figure 24. Updated filter mechanism test model.

This model includes four generic rules. The first is a traversal rule that checks the value of a fact in the link for crossing L1. The second is a rule that requires F6 to be true to set the value of F4 to true. The third is a rule that requires F4 to be true and F5 to be false to set the value of F6 to false. Finally, there is a rule that requires F6 to be false to set the value of F5 to true. These rules are shown in Table 2.

Table 2. Generic rules in the updated filter mechanism test model.

| ID | PREStart | POSTStart | PREEnd | POSTEnd | PRELink | POSTLink |
|---|---|---|---|---|---|---|
| 1 |  |  |  |  | P1:T | P1:T |
| 2 | P4:T |  | P2:F | P2:T |  |  |
| 3 | P2:T, P3:F |  | P4:T | P4:F |  |  |
| 4 | P4:F |  | P3:F | P3:T |  |  |

## 6. Experimentation Results Analysis

The scenarios selected from [2] were run using the multi-threaded SONARR algorithm and the resulting data was recorded. The results from the two threading implementations are discussed and compared in the following subsections.

### 6.1. Model 2

The two tests for model 2 demonstrated an increase in performance, which is shown in the data in Table 3. The other performance statistics for the tests remained the same.

Table 3. Single-threaded and multi-threaded comparison test for model 2.

| Scenario # | 3 | | 4 | |
|---|---|---|---|---|
| Threading | S | M | S | M |
| Time[†] | 00:01:33 | 00:00:52 | 00:00:13 | 00:00:09 |
| Final Paths | 123,919 | 123,919 | 4,056 | 4,056 |

[†] results do not include time taken to sort results

Scenario 3 started from container C2 and traversed to container C9. Rule R7 was omitted for this scenario, and fact F44 was set to true. The single-threaded algorithm took one minute and 33 seconds to complete and found 123,919 final reality paths. The multi-threaded algorithm took 52 seconds to complete and identified the same number of reality paths.

Scenario 4 started from container C2 and traversed to container C9. Rule R7 was also omitted for this test, and facts F37 and F38 were set to false. The single-threaded session took 13 seconds to complete and found 4,056 final reality paths. The multi-threaded session took nine seconds to complete and also found 4,056 final reality paths.

The difference in processing time between the two versions is small, as an absolute value. However, as a percentage it is notable. The multi-threaded algorithm took 59% of the time to complete scenario 3 and 69% of the time to complete processing scenario 4. This is a noticeable gain in performance and there was no loss in the quality of results. The larger of the two scenarios had a greater difference in completion time, both in terms of absolute difference and percentage. These results demonstrate the efficacy and benefit of the multi-threaded SONARR algorithm.

*6.2. Model 3*

The results from model 3, which are presented in Table 4, show that using the multi-threaded algorithm for scenarios 2 and 3 produced a notable performance gain, in terms of the completion time metric. Scenarios 4 and 5, scenarios with open parameters which allow for a large number of reality paths to be generated, produced different results due to the difference in rule omission before traversal. Nonetheless, both of the scenarios also show a gain in performance when utilizing the multi-threaded algorithm, as compared to the single-threaded algorithm.

Table 4. Single-threaded and multi-threaded comparison tests for model 3.

| Scenario # | 2 | | 3 | | 4 | | 5 | |
|---|---|---|---|---|---|---|---|---|
| Threading | S | M | S | M | S[††] | M | S[††] | M |
| Time[†] | 00:03:50 | 00:01:22 | 00:00:36 | 00:00:18 | 20:00:00 | 24:00:00 | 20:00:00 | 24:00:00 |
| Final Paths | 1,200,001 | 1,200,007 | 185,913 | 185,913 | 3 | 24,278,791 | 3 | 31 |

[†] results do not include time taken to sort results
[††] test was run on a debug build to monitor processing metrics

The traversal for scenario 2 started from container C22 and continued to container C29. Rule R1 was omitted, and fact F57 was set to false. In the single-threaded algorithm, this test was set to stop when the final reality path count exceeded 1.2 million. The multi-threaded test for this scenario was set to have the same termination condition to facilitate comparison between the two algorithms. The single-threaded algorithm took 3 minutes and 50 seconds to find 1.2 million reality paths. The multi-threaded algorithm took less than half that time, 1 minute and 22 seconds, to find the 1.2 million paths.

Since the memory limitations that impaired the operations of the single-threaded algorithm are not present in the multi-threaded implementation, a supplemental test was performed for scenario 2. This test aimed to determine whether the multi-threaded algorithm could complete the analysis (which the single-threaded algorithm could not, on the computer used for testing). The multi-threaded algorithm was run without the 1.2 million path termination condition. The final reality path count was calculated to be 9,864,101, in [2], based on the one-depth architecture of the subsection (as shown in Figure 25). The multi-threaded algorithm was able to complete processing and found all expected 9,864,101 reality paths in 9 minutes and 3 seconds. There was a total of 197,037,282 connections generated within these reality paths. The longest path recorded had a connection chain of 22 connections, which was the length of 3,626,801 reality paths. The shortest recorded path was two connections, with only one reality path of this length.

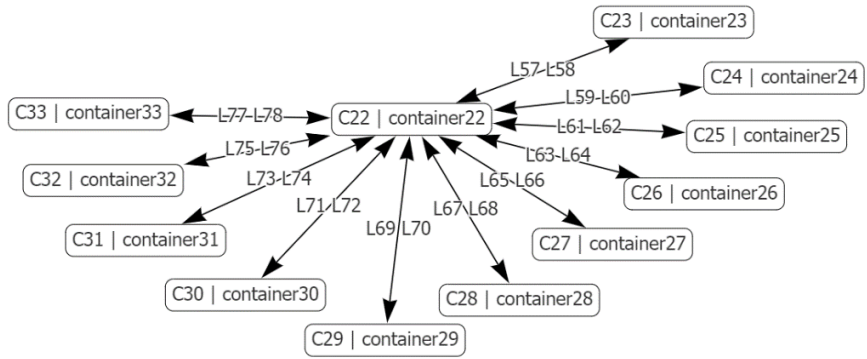

Figure 25. Subsection of model 3 used in scenario 2.

Scenario 3 started from container C1 and traversed to container C11. Rule R1 was omitted from this session, and three facts (F30, F33, and F38) were set to false to limit the traversal area to a specific subsection of the network. The single-threaded algorithm found 185,913 reality paths in 36 seconds. The multi-threaded algorithm also found 185,913 reality paths and only took 18 seconds. The multi-threaded algorithm was, thus, twice as fast for this scenario.

The single-threaded tests for scenarios 4 and 5 produced the same result of three reality paths, for each scenario, in 20 hours. The multi-threaded implementation tests produced unexpected results for these two scenarios.

The multi-threaded algorithm, when processing scenario 4, found 24,278,791 final reality paths from container C1 to container C15 in the 24-hour test window. The total number of connections generated for this session was 1.5 billion. There were 551,664 reality paths with the longest recorded connection chain length (of 70 connections) and only one reality path with the shortest chain (of 8 connections).

The multi-threaded testing for scenario 4 demonstrated the benefit of the multi-threaded algorithm. Seconds into processing, it had already identified more paths than the single-threaded algorithm. This was unexpected. It appears that the way the initial reality paths were created and distributed amongst the available processing threads allowed the multi-threaded algorithm to spread out rapidly, whereas the single-threaded algorithm processed paths one-by-one in a depth-first order. This initial branching effect, combined with generic rule R2, which prevents re-traversal across a link, allowed the multi-threaded algorithm to achieve superior results for scenario 4.

For scenario 5, the multi-threaded algorithm found 31 final reality paths from container C1 to container C15 in 24 hours. The total number of connections generated for these reality paths was 525. There were two reality paths with the longest recorded connection chain (24 connections), and one reality path with the shortest chain (8 connections).

Scenario 5's results were quite different to scenario 4's. The difference in parameters between the two scenarios is that scenario 5 had a rule that did not prevent re-traversal, while the rule used in scenario 4 did prevent re-traversal. Both scenarios started with the same initial branching effect, described above; however, scenario 5 presented significantly more paths for the algorithm to process, due to not having the re-traversal prevention rule limitation. This demonstrates how the network and the rule-fact logic within the network plays a significant role in the operations and performance of SONARR. The multi-threaded algorithm has been demonstrated to have a notable improvement in performance, as compared to the original single-threaded algorithm; however, it does not eliminate the need to ensure that the rule-fact logic is designed to promote efficiency, to the greatest extent possible, given testing scenario requirements.

*6.3. Completion filter tests*

This section presents the results of testing of the completion filter mechanism. Each scenario that was tested had different completion filter parameters set before traversal. Scenario 1 had no filters applied. Scenario 2 had only a F4 must be true filter applied to C2. Scenario 3 had a F4 and F5 must be true filter applied to C2. Finally, scenario 4 had a F4 or F5 must be true filter applied. The results of the scenarios are shown in Table 5.

Table 5. Results of the updated filter mechanism tests.

| Scenario # | 1 | 2 | 3 | 4 |
|---|---|---|---|---|
| Start Container | C1 | C1 | C1 | C1 |
| End Container | C2 | C2 | C2 | C2 |
| Applied Filters | --- | F4:T | F4:T and F5:T | F4:T or F5:T |
| Completion Time | Immediate | Immediate | Immediate | Immediate |
| Connections | 2 | 4 | 6 | 4 |
| Rules Triggered | 1 | 4 | 8 | 4 |

Scenario 1 provides a baseline for comparison. It shows that, without any filters applied, SONARR will stop its traversal at the specified end container. This model has one link, from C1 to C2, so the algorithm traverses that link and does not go further because it has satisfied its goal.

Scenario 2 illustrates the use of a filter to traverse beyond a specified end container. The F4 must be true completion filter used in this scenario means that F4, which is stored in the specified end container C2, must have a value of true for a reality path to complete. To satisfy this filter, SONARR had to traverse past C2 and change the value of F6 in C3 to trigger a rule that sets F4 in C2 to have the value of true. This satisfies the filter and, thus, the traversal ends.

Scenario 3 used a filter that required F4 and F5 to be true, illustrating the usage of the "and" propositional logic in filters. This traversal required that SONARR pass through the end container twice, before ultimately satisfying the filter. Continuing from the path discussed for scenario 2, the algorithm traverses to C3, setting the value of F6 to false, and then back to C2, setting the value of F5 to true. At this point both F4 and F5 have the value of true, satisfying the filter and terminating traversal.

Scenario 4 illustrates usage of "or" propositional logic in filters. Because the filter specifies either F4 or F5 needing to be true, the first occurrence of one of them, which happens to be F4, being set to true satisfies the filter and ends the traversal. The results of this scenario are the same as for scenario 2, as F4 having a value of true is all that was necessary to end the traversal.

This testing demonstrated the efficacy of the completion filter mechanism and demonstrated its impact on traversal processing.

### 7. Conclusions and Future Work

As expected, the multi-threaded algorithm increased the performance and output, in a given period of time, of the SONARR software. The multi-threaded algorithm was shown to provide scalable performance, directly correlating with the number of processors available. Since the algorithm dynamically allocates threads based on the number of processors available, the utilization of multiple CPUs will facilitate the processing of large, complex SONARR networks.

The storage capacity of the computer running the SONARR analysis could still pose a limitation. If the results of a session get too large, the computer may have insufficient storage space. The file-writing approach used by the multi-threaded algorithm eliminates the limitation of available memory for processing and replaces it with the limitation of available disk space. Large capacity hard disks can be used to mitigate this limitation.

The updated completion filtering mechanism allows users to precisely specify the desired goal state of the end container. It implements as propositional logic and provides the algorithm with another termination control. When used appropriately, the filter can significantly reduce the number of resulting reality path for users to analyze.

Future work will focus on continuing to make the exhaustive SONARR algorithm faster. One planned area of work is to use a tree of connection objects, instead of a list of reality paths during traversal. The connection object is currently used in the reality path to create a history that specific path took through the network, storing the fact changes and rules triggered. The algorithm constantly makes new connection objects and adds them to the reality path being traversed. Because these connections may be traversed and recreated in different reality paths, this results in SONARR performing redundant work during processing. Instead, connection objects can potentially be reused by other reality paths.

This approach is designed to prevent duplicate work past a connection. If a traversed connection has already been created, and all paths branching from it have been fully traversed, any reality path that reaches that connection and has the same network configuration as that connection would have the exact same results past that connection, regardless of the path taken to get there. Thus, there is no need to perform these traversals again, reducing the work performed by SONARR. To support this, the algorithm will need to be modified to accommodate new objectives in the storage and retrieval of traversed node information. Several core objects will also have to be altered to hold new property information.

Another area of planned future work is testing the performance of the multi-threaded SONARR algorithm on a CPU cluster. This work will seek to characterize the level of performance improvement that an environment with a greater number of CPUs provides.

The final planned area of future work is augmenting the controls and heuristics used in the SONARR algorithm. The current controls help prevent processing unnecessary paths. It is also desirable to add a capability to determine the best routes through the network. A common heuristic value, that can be used on links, which will increment during traversal and signify the link's attractiveness to reaching an end goal will be evaluated. This is expected to help classify paths and identify the most important and plausible attack paths.

## Acknowledgements

Thanks is given to Anthony DeFoe for his work on visualization development for the software. Thanks is also given to other members of the project team for their feedback, testing and other contributions.